\documentstyle[aps,preprint]{revtex}
\begin{document}
\draft
\title{\large
Chaotic enhancement of hydrogen atoms excitation in magnetic
and microwave fields} 
\author{Giuliano Benenti, Giulio Casati}
\address{Universit\`a di Milano, sede di Como, Via Lucini 3,
22100 Como, Italy}
\address{Istituto Nazionale di Fisica della Materia, 
Unit\`a di Milano, Via Celoria 16, 20133 Milano, Italy}
\address{Istituto Nazionale di Fisica Nucleare, Sezione di Milano,
Via Celoria 16, 20133 Milano, Italy} 
\author{Dima L. Shepelyansky$^{*}$}
\address{Laboratoire de Physique Quantique, UMR C5626 du CNRS, \\
Universit\'e Paul Sabatier, 31062,
Toulouse, France}
\date{29 December 1996}
\maketitle
\begin{abstract}
We numerically investigate multiphoton ionization of excited 
hydrogen atoms in magnetic and microwave fields when up to 
\hbox{$N_I=600$}
photons are required for ionization. The analytical estimates for
the quantum 
localization length in the classically chaotic regime are in agreement 
with numerical data. The excitation is much stronger as compared
to the case with microwave field only due to the chaotic structure 
of eigenstates in magnetic field. 
\end{abstract}

In 1982 Shushkov and Flambaum \cite{ref1} discussed the effect of
weak interaction enhancement due to a complex structure of
ergodic eigenfunctions in nuclei. The basic idea of this effect is 
that in complex systems an eigenfunction, represented in some basis, 
has a large number $M$ of randomly fluctuating components so that their
typical value is $1/\sqrt{M}$. Due to this, the matrix elements for
interparticle interaction are $V_{int} \sim 1/\sqrt{M}$ 
while the distance between mixed levels is $\Delta E\sim 1/M$. 
As a result, according to the perturbation theory, the admixture
factor $\eta$
is strongly enhanced: $\eta\sim V_{int}/\Delta E\sim \sqrt{M}$
as compared to the case in which eigenfunctions have only few
components $(M \sim 1)$. 
This effect was investigated and well confirmed in experiments with
weak interaction enhancement for scattering of polarized 
neutrons on nuclei \cite{ref1}.
Recently a similar effect of interparticle interaction enhancement
was discussed for two interacting particles in disordered solid
state systems \cite{ref2}. Here, a short range interaction produces
a strong enhancement of the localization length leading to a qualitative
change of physical properties. This shows that the effect is quite
general and can take place in different systems.

In this Letter we investigate the possibility of similar
enhancement in atomic physics for atoms interacting with
electromagnetic fields. Such process becomes especially interesting
for highly excited atoms (hydrogen or Rydberg atoms) in microwave
fields where absorption of many photons is necessary in order to
ionize electrons.
Until now this problem was studied only in the case in which
the electron dynamics, in absence of microwave field, is
integrable \cite{ref3}.
In this case strong ionization is possible due to onset of chaos 
at sufficiently strong field intensity. As it is known, above the
classical chaos border ionization proceeds in a diffusive way and 
quantum interference effects can lead to localization of this diffusion
and thus suppress ionization.

A quite different situation, which 
was never studied neither numerically nor experimentally,
appears when the electron's motion in the atom is 
already chaotic in the absence of microwave field. 
An interesting example of such 
situation is an hydrogen atom in a strong static magnetic field.
The properties of such atoms have been extensively studied in the last
decade \cite{ref4,ref5} and it has been shown that the eigenfunctions
are chaotic, and that several properties of the system can be described by 
Random  Matrix Theory.
Due to that one can expect that the interaction of such an atom with
a microwave field will be strongly enhanced so that the localization length 
will become much larger than the corresponding one in the absence of magnetic
fields. As a result, the quantum delocalization border, which determines
the ionization threshold, will be strongly decreased. 

The investigation of classical dynamics and some preliminary estimates
for the quantum localization length in such a case 
have been given in a recent paper \cite{ref6}.
Here we discuss the quantum dynamics and present the results of
numerical simulations which confirm the chaotic enhancement
of quantum excitation as compared to the case without magnetic
field. 
Our studies also show a number of interesting features which
arise in this model in the adiabatic regime when the microwave
frequency is much smaller than the Kepler frequency. 

We consider the case in which the electric and magnetic fields are
parallel. In this case the magnetic quantum number $m$ is an exact 
integral of motion and here we set $m=0$.
The Hamiltonian writes
\begin{equation}
H = {{p_z^2}\over{2}}+{{p_\rho^2}\over{2}}+
{{\omega_L^2 \rho^2}\over{8}}-{1\over\sqrt{z^2+\rho^2}}+
\epsilon z \cos (\omega t)\,,
\label{eq1}
\end{equation}
where $\omega_L = B/c = B\hbox{(T)}/B_0$ is the
cyclotron frequency, 
$B_0 = 2.35 \times 10^5 \hbox{T}$, 
$\epsilon$ and $\omega$ are the field 
strength and frequency respectively (atomic units are used).
As it is known \cite{ref4,ref5}, in the absence of microwave field,
the classical motion becomes chaotic for $\omega_L n_0^3 \agt 1$ and
no visible islands of stability are present for $\omega_L n_0^3 
\approx 9$.  For $\omega_L n_0^3=3$ some islands of stability
exist but their size is small.

The turn on of microwave field leads to diffusive energy growth with
classical diffusion rate per unit time $D_B=(\Delta E)^2/\Delta t$.
The dependence of $D_B$ on parameters $\epsilon$, $\omega$ has been 
found in \cite{ref6}:
$D_B/D_0\approx \chi_1\omega_0^2$ $(\omega_0\ll 1)$,
$D_B/D_0\approx \chi_2/\omega_0^{4/3}$ $(\omega_0\gg 1)$,
where $\omega_0=\omega n_0^3$,
$D_0=\epsilon^2 n_0/2$ is the diffusion rate in the chaotic regime
for $B=0$ and $\omega_0=1$ and $\chi_1$, $\chi_2$ are 
two constants, weakly dependent on magnetic field (numerically, 
$\chi_1 \approx 18$, $\chi_2 \approx 2$ at $\omega_L n_0^3=9.2$
and $\chi_1 \approx 25$, $\chi_2 \approx 1$ at $\omega_L n_0^3=3$).
The above estimates for $D_B$ give the asymptotic behavior
of the diffusion rate for very small and very large $\omega_0$,
while the actual values of $D_B$ were determined from numerical 
simulations of the classical problem.

In the quantum case the interference effects can lead to localization 
of this diffusion \cite{ref6,ref7}. The localization length in
number of photons $\ell_B$ is proportional to the one--photon transition
rate $\Gamma$ and to the density of states $\rho_B$  
coupled by these 
transitions \cite{ref8}: $\ell_B \sim \Gamma\rho_B$.
The transition rate $\Gamma$ can be derived from the classical diffusion 
rate: $\Gamma\approx D_B/\omega^2$. We recall that for the case $B=0$ the
localization length $\ell_\phi$ at $\omega_0=1$ is $\ell_\phi=
3.3\epsilon_0^2 n_0^2 \sim D_0 n_0^6 \rho_0$ \cite{ref7}, where
$\rho_0=n_0^3$ is the density of states and $\epsilon_0=\epsilon n_0^4$.
Due to Coulomb degeneracy and to the existence of an additional 
approximate integral of motion \cite{ref7} the density $\rho_0$ is $n_0$
times smaller than the number of levels in one unit energy interval.
As a result \cite{ref6} 
\begin{equation}
\ell_B=\ell_\phi \,\frac{D_B}{D_0\omega_0^2}\,
\frac{\rho_B}{\rho_0}\,,
\label{eq2}
\end{equation}
where it is assumed $\ell_B > 1$ and $\omega\rho_B > 1$.
According to our quantum data  $\rho_B/\rho_0\approx
n_0/(\omega_L n_0^3)$ for $\omega_L n_0^3 > 1$.
More exactly $\rho_B/\rho_0=0.34 n_0$ (for $\omega_L n_0^3=3$, $n_0=60$)
and $\rho_B/\rho_0=0.14 n_0$ (for $\omega_L n_0^3=9.2$, $n_0=60$).
The dependence $\rho_B\sim 1/\omega_L$ is due to the oscillatory 
type behavior in $\rho$ direction in (\ref{eq1}).
The number of photons required for ionization is $N_I=n_0/2\omega_0$
and therefore for $\ell_B\ll N_I$ eigenfunctions are exponentially localized
in the number of photons $N_\phi$, namely 
$\psi_N\sim \exp(-|N_\phi|/\ell_B)$.

The value of the localization length $\ell_B$ is strongly enhanced
compared to the length $\ell_{\phi_\omega}=3.3 \epsilon_0^2 n_0^2/
\omega_0^{10/3}$ at $B=0$ and $\omega_0 > 1$. 
The enhancement factor $\ell_B/\ell_{\phi_\omega}
\approx  \chi_2 \rho_B/\rho_0 \approx 
\chi_2 n_0/(\omega_L n_0^3)\gg1$ is proportional
to the initially excited state $n_0$. In fact in the presence of a 
magnetic field
there is no additional integral of motion \cite{ref7} and the number 
of components in the eigenfunctions is increased by a factor 
$M=\rho_B/\rho_0$.  As a result the admixture factor $\eta$ is also
enhanced, namely $\eta^2\sim M$ similarly to the enhancement of localization
length in disordered solid state models with two particles \cite{ref2}
($\ell_B/\ell_{\phi_\omega} \sim \eta^2\sim M$).

The condition  $\ell_B=N_I$ gives the delocalization border 
$\epsilon_q$ above
which quantum excitation is close to the classical one
(both for $\omega_0 < 1$ and $\omega_0 \geq 1$):
\begin{equation}
\epsilon_0 > \epsilon_q=\frac{1}{n_0}\,\sqrt{\frac{D_0\,\omega_0^2}
{6.6\, D_B} \,\frac{\rho_0\, n_0}{\rho_B\,\omega_0}}.
\label{eq3}
\end{equation}
For $\omega_0 \geq 1$
this value is approximately by the factor $(3/n_0)^{1/2}$
below the delocalization border in microwave field 
only ($B=0$) where $\epsilon_{q0} \approx \omega_0^{7/6}/\sqrt{6.6 n_0}$.
For $\omega_0 \ll 1$ the boder is $\epsilon_q =
(\omega_L/6.6\chi_1\omega)^{1/2}/n_0$.

In order to check the above estimates (eqs. (\ref{eq2}) and (\ref{eq3})) 
we analyzed the quantum dynamics following the wave packet evolution in 
the eigenstate basis at $\epsilon=0$. Initially only one eigenstate
is excited with eigenenergy $E_{\lambda_0}\approx E_0=-1/2n_0^2$ and 
$n_0=60$. In our computations we used a
total number of eigenstates up to $800$ and the
evolution was followed up to time $\tau=200$ 
microwave periods. The parameters were varied in the intervals $0.05\leq
\omega_0\leq 3$, $0.002\leq\epsilon_0\leq 0.02$ for $\omega_L n_0^3=3$
and $9.2$.
For this parameter range, the number of photons $N_I=n_0/2\omega_0$
required for ionization varies in the interval $10\leq N_I\leq 600$.
The probability distribution $f_\lambda$ over the eigenstates
at $\epsilon=0$ is shown in figs.~\ref{fig1},~\ref{fig2} as a function 
of the number of absorbed photons
$N_\phi=(E_\lambda-E_0)/\omega$.
In order to suppress fluctuations this probability was averaged over
$10-20$ microwave periods. For the comparison with classical results
we also determined the probability  $f_N$ in each one--photon interval
around integer values of $N_\phi$. The classical distribution was obtained
by solution of Newton equations with up to $5 \times 10^3$ 
classical trajectories 
and was normalized to one--photon interval. Initially the trajectories
were distributed microcanonically on the energy surface at energy
$E_0$.

The typical results in the localized regime are presented in 
fig.~\ref{fig1}. Here the distribution reaches its stationary state 
with a well localized exponential profile. The least square fit
with $f_N \sim \exp(-2 N_\phi/\ell_{BN})$ for $N_\phi \geq 0$ 
allows to determine the numerical 
value of the localization length $\ell_{BN}$ which turns out to be
in good agreement with
theoretical estimate (eq. (\ref{eq2})) and is strikingly enhanced compared
to the case of zero magnetic field.
The plateau which appears in fig.~\ref{fig1}b for $N_\phi > 130$ is
related to the finite  size of the basis and to the fact that, according
to eq. (\ref{eq2}), the localization length is non homogeneous on high
levels ($\ell_B \sim n_0^{11}\sim \left(N_I-N_\phi\right)^{-11/2}$ for
$\omega_0 \ll 1$).

In fig.~\ref{fig2} the distributions in the delocalized case are 
shown for $\omega_0=1$ (fig.~\ref{fig2}a) and $\omega_0=0.1$
(fig.~\ref{fig2}b). The delocalization borders in these cases,
$\epsilon_q=0.016$ (fig.~\ref{fig2}a) and $\epsilon_q=0.014$,
(fig.~\ref{fig2}b) are below the field peak value $\epsilon_0=0.02$.
The numerical results show a good agreement between classical
and quantum distributions in this regime. 
Even if at $\epsilon_0=0.02$ the dynamics starts to be 
chaotic also at zero magnetic field, the 
excitation at $\omega_L n_0^3=3$ is much
stronger due to the chaotic enhancement of electron's interaction 
with the microwave 
field. For example, the increase of 
$\left<(\Delta N_{\phi})^2\right>$ after 50 
microwave periods is approximately 55 for the case of fig.~\ref{fig2}a
while at $B=0$ it is only 6.

In order to check the theoretical predictions for the photonic localization
length $\ell_B$ we analyzed different probability distributions in the
localized regime for $0.05\leq \omega_0\leq 3$ at $\omega_L n_0^3=3$
and $\omega_L n_0^3=9.2$. The comparison of the numerically obtained
lengths $\ell_{BN}$ with the theoretical estimate eq. (\ref{eq2}) is
presented in fig.~\ref{fig3}. Without any fitting parameter the
theoretical line demonstrates the fairly good agreement with numerical data.
Indeed the average value of the ratio $R=\left< \ell_{BN}/
\ell_B\right>=0.81\pm 0.34$, where the average is over all values 
$\ell_B>1$. The separate averaging over the cases with $\omega_0
\geq 1$ and $\omega_0<1$ gives $R_1=0.98\pm 0.30$ and $R_2=0.70\pm 0.26$
respectively.

In spite of the good agreement between theoretical predictions and
numerical data we would like to stress that 
a deeper investigation of
the problem is required. Especially unusual is the regime
$\omega_0<1$, which has not been studied up to now from the
viewpoint of dynamical localization. A number of new questions 
appear in this regime. For example for $\omega_0>1$ a chain of
one--photon transitions is clearly seen in the quantum localized
distribution (fig.~\ref{fig1}a) while for $\omega_0<1$ the 
structure is not visible even though $\omega \rho_B >1$
(fig.~\ref{fig1}b).

Another interesting question in this adiabatic regime $\omega_0\ll 1$
is connected with the possibility of analyzing the problem 
in the instantaneous time basis. In this basis the Hamiltonian
takes the form $H(t)=H_0(t)+\partial S/\partial t $, where
$H_0$ is the Hamiltonian (\ref{eq1}) at a given moment of time while
$\partial S/\partial t $ describes the transitions due to the field's
variation with time ($S$ is the action of the Hamiltonian (\ref{eq1})
in which time is considered as a parameter). The term
$\partial S/\partial t$ can be estimated as 
$\partial S/\partial t\sim \epsilon\,\omega\sin (\omega t)
\,2\pi n^3 z \sim \epsilon \,\omega\, n^5$ (see also \cite{ref9}).
It describes the transitions between instant time levels of the Hamiltonian
(\ref{eq1}), the amplitude of which can be estimated as $V_{eff}\sim
\epsilon\omega n^5/\sqrt{n}$. The factor $\sqrt{n}$ in the
denominator appears due to the chaotic structure of eigenstates 
which leads to a smearing of $\partial S/\partial t$
over the $n$ states which contribute to the eigenfunctions inside
an atomic shell (we assume $\omega_L\sim 1/n_0^3\,$). Since the distance
between levels is $\delta E\sim 1/\rho_B\sim n_0^{-4}$ it seems 
that mixing between instant levels is possible only if $V_{eff}>
\delta E$, giving $\epsilon_0 \,\omega_0 \,n_0^{3/2} > 1$. This adiabatic
condition is more restrictive than the standard $\ell_B>1$
($\epsilon_0 \,n_0^{3/2} > 1$). However the numerical results
(fig.~\ref{fig3}) confirm our estimate (\ref{eq2}). For a
possible explanation of this discrepancy one may argue that 
the distance between coupled quasi--energy levels
is $\delta E_\omega \sim \omega/n$ and then the condition
$V_{eff}>\delta E_\omega$
gives $\ell_B>1$ in agreement with the estimate (\ref{eq2}). 
Another possible
reason is that in the instant time basis the levels are moving with
time and can therefore intersect each other giving $\delta E=0$.

In conclusion our numerical investigations confirm the theoretical
estimates for the photonic localization length (\ref{eq2}) both
for $\omega_0 \geq 1$ and $\omega_0<1$. Due to the chaotic structure of
the eigenstates the quantum delocalization
border is strongly lowered compared to the case with microwave 
field only. Since for $\omega_0\ll 1$ a much larger number of
photons is required for ionization ($N_I=300$ in figs.~\ref{fig1}b,
\ref{fig2}b) experimental observation of localization and verification
of theoretical predictions should be more easily 
feasible in laboratory experiments.

\begin{figure}
\caption{
Probability distribution as a function of photon number $N_\phi$.
Quantum distribution $f_\lambda$ over the eigenbasis at $\epsilon=0$
(full line); quantum probability in one--photon interval $f_N$
(circles); classical distribution in one--photon interval (dashed line).
The straight line shows the fit for the exponential decay.
\protect\newline
(a) $n_0=60$, $\omega_0=1$, $\omega_L n_0^3=3$, $\epsilon_0=0.005$,
$\ell_{BN}=3.6$, $\ell_B=3$, $180\leq\tau\leq 200$, $D_B/D_0=0.49$.
\protect\newline
(b) $n_0=60$, $\omega_0=0.1$, $\omega_L n_0^3=3$, $\epsilon_0=0.005$,
$\ell_{BN}=29.5$, $\ell_B=37.5$, $180\leq\tau\leq 200$, $D_B/D_0=0.062$.
}
\label{fig1}
\end{figure}

\begin{figure}
\caption{
Same as in fig.~\ref{fig1} but in the delocalized regime, with 
$\epsilon_0=0.02 > \epsilon_q\approx 0.015$, and $40\leq\tau\leq 50$.
The classical (dashed line) and quantum (circles) distributions in 
one--photon interval are close to each other.
}
\label{fig2}
\end{figure}

\begin{figure}
\caption{
The numerically computed localization length $\ell_{BN}$ 
versus the theoretical
estimate $\ell_B$ eq. (\ref{eq2}) for $\omega_L n_0^3=3$,
$\omega_0\geq 1$ (full circles), $\omega_L n_0^3=3$,
$\omega_0 < 1$ (full triangles), $\omega_L n_0^3=9.2$,
$\omega_0\geq 1$ (open circles), $\omega_L n_0^3=9.2$,
$\omega_0 < 1$ (open triangles). The error bars obtained 
from least square fits of the localized distributions are also shown.
The straight line corresponds to
$\ell_{BN}=\ell_B$.
}
\label{fig3}
\end{figure}

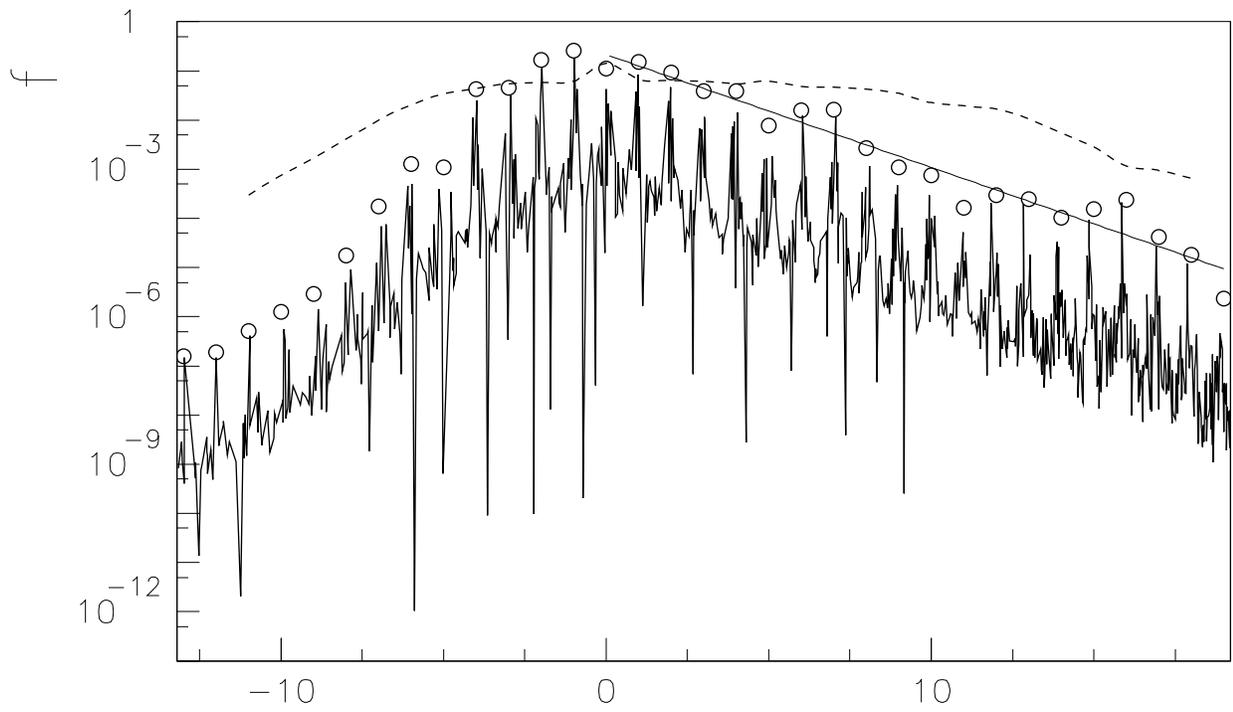
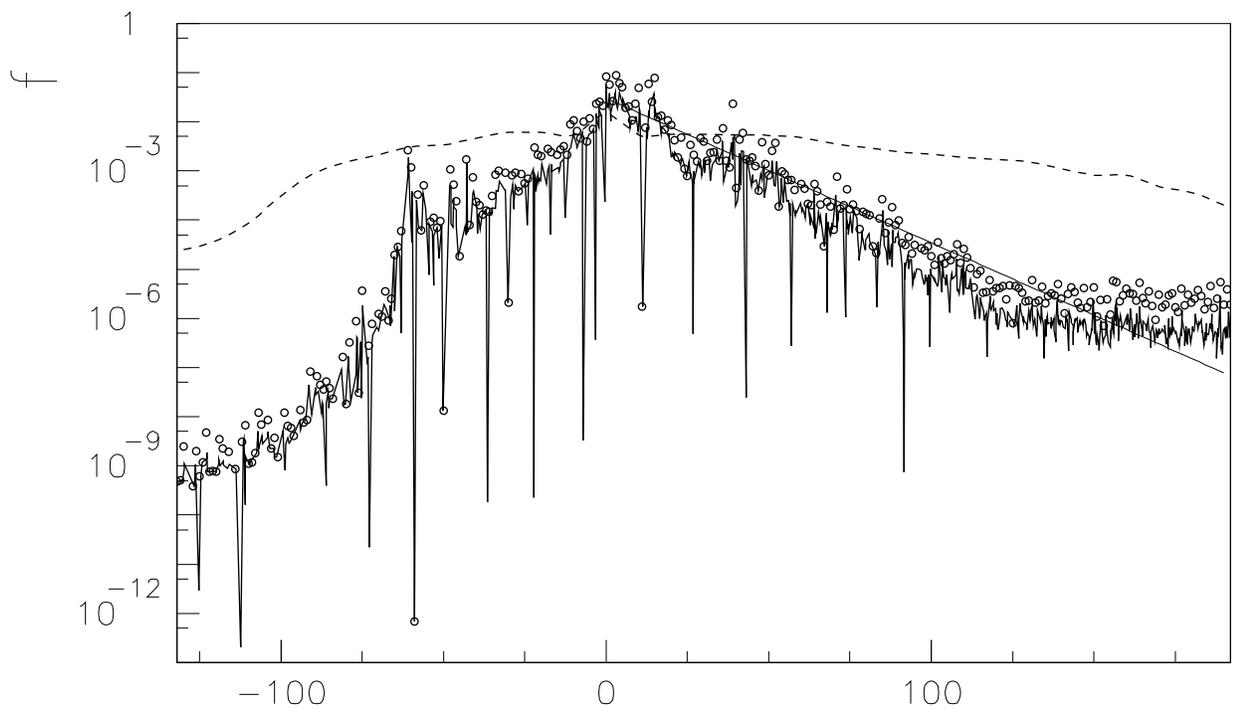
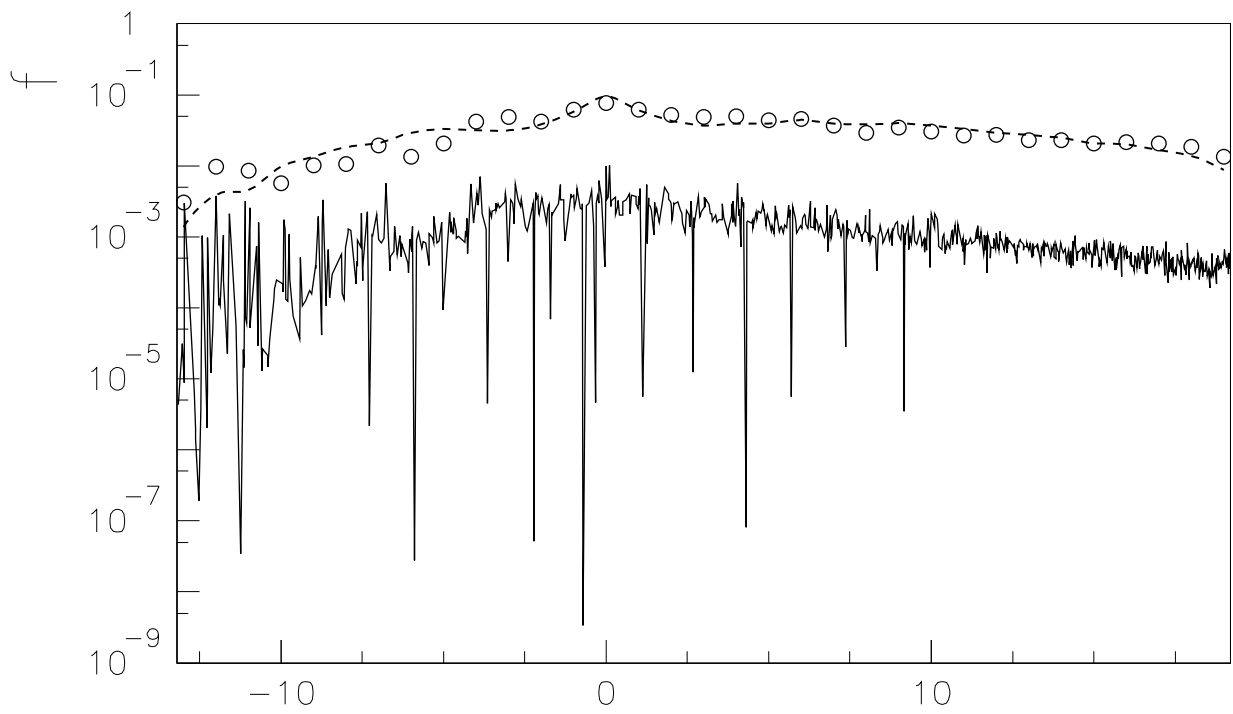
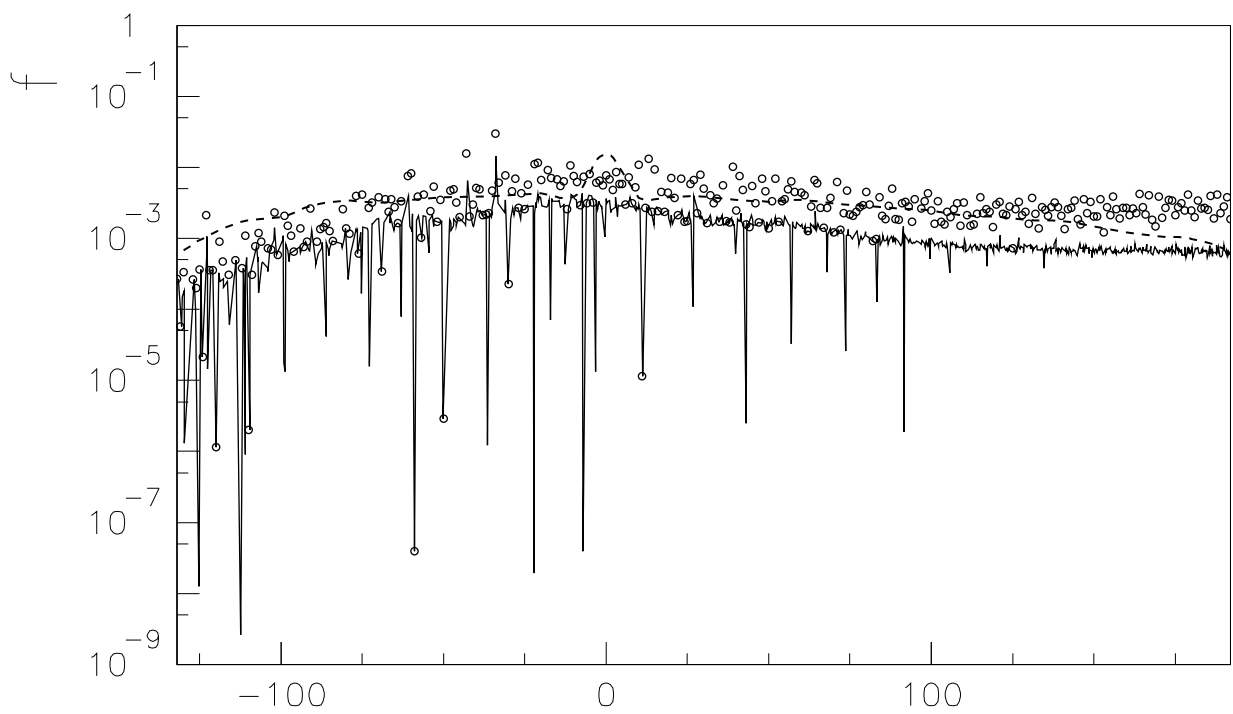
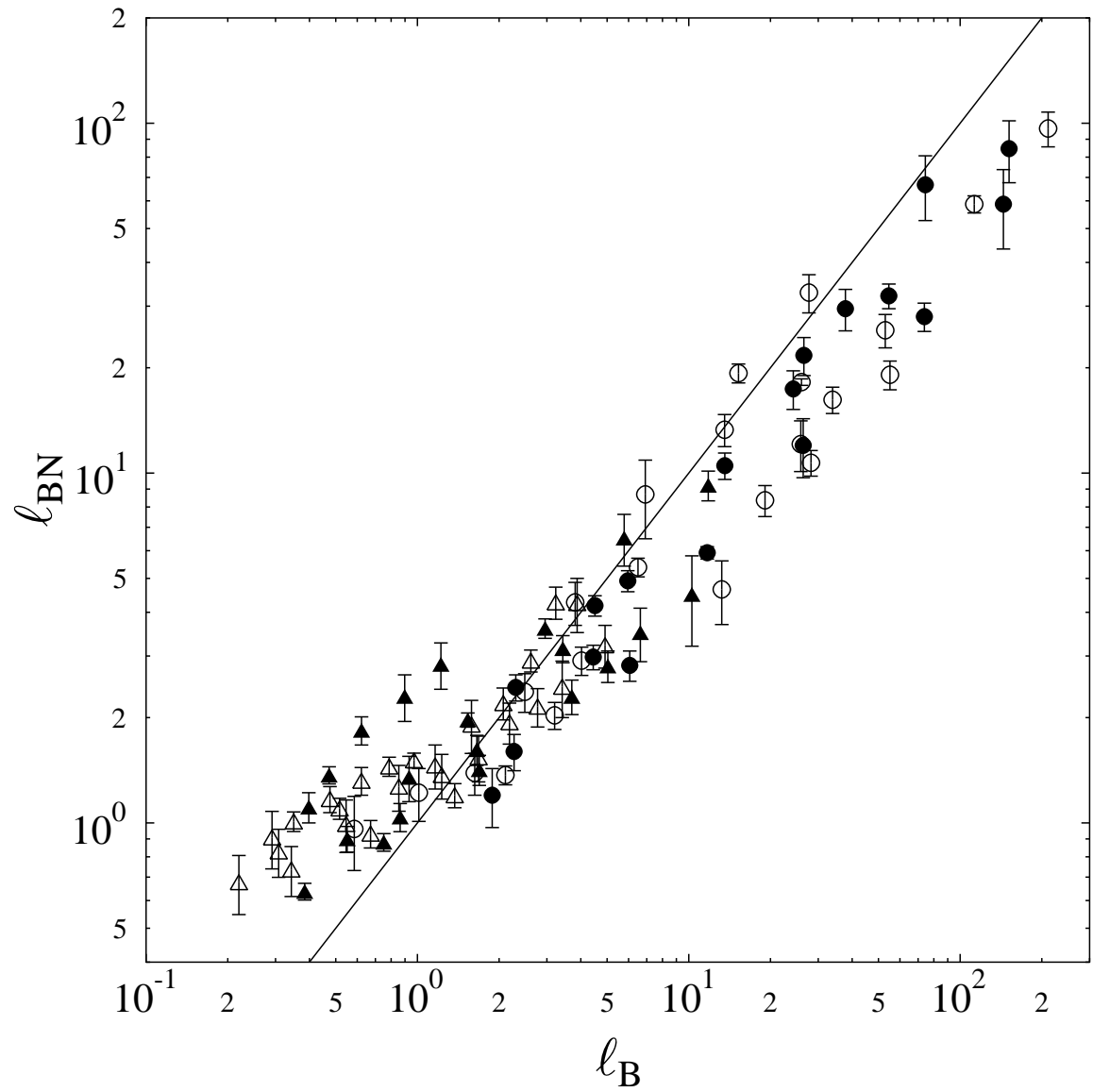
\end{document}